\documentclass[a4paper]{jpconf}
\usepackage{graphicx}
\usepackage[centertags]{amsmath}
\usepackage{amsfonts}
\usepackage{amssymb}
\usepackage{amsthm}
\usepackage{newlfont}
\usepackage{stmaryrd}
\usepackage{mathrsfs}
\usepackage{euscript}
\usepackage{graphicx}
\usepackage{enumerate}
 \usepackage[final]{changes}
 
\usepackage{wrapfig}
\usepackage{amscd}
\usepackage{hyperref}

\usepackage{caption}
\usepackage{subcaption}

\theoremstyle{plain}

\theoremstyle{definition}

\theoremstyle{remark}

\newcommand{\id}{\textrm{d}}
\newcommand{\opunit}{\text{1}\kern-0.22em\text{l}}

\DeclareMathAlphabet{\mathpzc}{OT1}{pzc}{m}{it}

\begin{document}

\title{Nonequilibrium Third Law of Thermodynamics}

\author{Faezeh Khodabandehlou}
\ead{faezeh.khodabandehlou@kuleuven.be}
\address{Department of Physics and Astronomy, KU Leuven, Belgium}

\begin{abstract}
The extended Third Law of thermodynamics for nonequilibrium jump processes is studied in previous collaborative papers (Kh et al.,2023) \cite{nernst, SM}, where the nonequilibrium heat capacity and the excess heat with the corresponding quasipotential are introduced. The extended Third Law states that the nonequilibrium heat capacity vanishes under two conditions as the temperature approaches zero. The current paper presents a concise overview of key papers addressing the nonequilibrium calorimetry and the nonequilibrium Nernst postulate (the nonequilibrium Third Law of thermodynamics). The quasipotential is a crucial quantity to calculate the nonequilibrium heat capacity. The new result presented in this work includes an interpretation of the quasipotential in terms of the mean time taken by the Markov jump process, starting from a given state, to reach other states for the first time.
\end{abstract}

\section{Introduction}
The Third Law of Thermodynamics, initially formulated by Nernst and Planck, examines fundamental questions about heat capacity and entropy at low temperatures \cite{ner,pla}. Nernst's version suggests that entropy becomes a constant at zero temperature, independent of external parameters, and that heat capacity vanishes as the temperature approaches zero \cite{cal, mas}. However, the Third Law or the Nernst postulate is not universally applicable to all substances and requires nontrivial conditions \cite{lie1,lie2}. \\
Over time, the emphasis has been on developing calorimetry and establishing a general structure for studying thermodynamic processes in nonequilibrium systems \cite{pri,oon,kom2}. The objective is not solely to extend the formal definition of entropy. Instead, it aims to obtain quantitative measures of heat changes during quasistatic relaxations in small open systems under new nonequilibrium conditions. If the change is induced by modifying the temperature of the environment, the excess heat per change in temperature is appropriately termed nonequilibrium heat capacity \cite{eu,jir,cal2}. This need not be restricted to the context of linear irreversible thermodynamics or close-to-equilibrium processes. We are interested in the low-temperature behaviour of excess heat resulting from changes in system parameters during such quasistatic relaxations \cite{pri,oon,kom2, st,clau}.\\

Our main result indicates that the excess heat vanishes at absolute zero for a large and physically motivated class of nonequilibrium systems modelled as Markov jump processes. Similar to its equilibrium version, this result requires two conditions for extending the Third Law: the uniqueness of the ground state and the sufficient accessibility of states to each other. For the second condition, we will show that the maximum mean time for the random walker starting from any state to reach different states must remain finite at vanishing temperatures.

\section{Setup}
Nonequilibrium calorimetry, a relatively new subject in nonequilibrium statistical physics, has been explored in other works; see \cite{cal2, pri}. However, in this paper, our focus is on discussing the nonequilibrium heat capacity at low temperatures. The Markov jump process is the appropriate and reasonable model for describing low-temperature thermal properties. We denote the process at time $t$ as $X_t$.\\
Consider a finite set of physical states represented by variables like $x, y, z, \ldots$, each corresponding to energy levels $E(x)$; for instance, those linked to molecules or lattice gases. In the Markov jump process, transitions occur from state $x$ to $y$ with a transition rate denoted as $k(x,y)$. Each state can be assigned a vertex, and an edge exists between $x$ and $y$ only if a possible jump exists between these two states. As a result, we have a simple finite graph. Let $L$ be the backward generator  of the Markov jump process, and it operates as follows:
\begin{equation}
   L\, h(x)=\sum_y k(x,y) (h(y)-h(x)), \qquad e^{tL}h\, (x)=\langle  h (X_t)|X_0=x\rangle,
\end{equation}
for any function of $h$. The transition rates contain kinetic and thermodynamic details, summarizing the weak coupling of the system to a thermal bath. Typically, these rates may depend on a multi-dimensional vector in parameter space; see \cite{SM}. The rates are taken to satisfy the local detailed balance, 
\begin{equation}\label{ldb}
\log\frac{k_\beta (x,y)}{k_\beta(y,x)} = \beta\, q(x,y),
\end{equation}
where the antisymmetric $q(x,y)=-q(y,x)$ is the heat dissipated in the thermal bath at inverse temperature $\beta$; see \cite{ldb} and references therein. \\
The rates appear in the master equation, which, in the stationary case, is expressed as:
\begin{equation}
    \sum_y k_\beta(y,x)\rho_\beta(y)-k_\beta(x,y)\rho_\beta(x)=0,
\end{equation}
$\rho_\beta(x)$ is the final solution and represents the unique stationary distribution of state $x$. The expectation of the dissipated power or heat flux into the thermal bath is 
\begin{equation}\label{ju}
\cal P_{\beta}(x) = \sum_y k_{\beta}(x,y) \, q(x,y),
\end{equation}
when in state $x$. The stationary average is denoted by  $ \langle \cal P_{\beta} \rangle_{\beta}=\sum_x \cal P_{\beta}(x) \rho_\beta(x) \geq 0$, where positive values signifies a nonequilibrium condition.

\section{Excess heat and heat capacity}
Suppose the initial temperature of the thermal bath at time $t=0$ is $\beta$. If we perturb $\beta$ slightly, $\beta'=\beta +\id \beta$, the system relaxes from state $x$ to the new steady state, exchanging new excess heat with the thermal bath. We define the quasipotential as:
\begin{equation}\label{psp}
V_{\beta'}(x) := \int_0^\infty\id t\,[\langle \cal P_{\beta'}(X_t\,|\,X_0=x) \rangle_{\beta'} - \langle \cal P_{\beta'} \rangle_{\beta'}],
\end{equation}
where expectations refer to relaxation, resp. stationarity under the nonequilibrium process for control $\beta'$.
% From \eqref{kt}, the expected excess work on the system when starting in $x$ is given by the pseudo-potential
%where the expectations $\langle \cdot \rangle'$ use the new value $\beta'$ to run the dynamics, and $\langle w\rangle'$ is the new stationary dissipated power.  %Note that $\langle V \rangle =0$.  
The time-integral converges by well-known results on finite-state Markov processes. The excess heat towards the system is then $Q^\text{exc}_{\beta'}(x) = - V_{\beta'}(x)$, and we must draw $x$  from the initial stationary distribution at the original $\beta=\beta' - \id \beta$ .  Taking that zero-time expectation over $x$, we find
\begin{equation}
\langle Q^\text{exc}_{\beta'}\rangle_\beta :=  \langle V_\beta\rangle_\beta - \langle V_{\beta'}\rangle_\beta, 
\end{equation}
on the other hand,   $\langle V_\beta\rangle_\beta =0$. Thus, it follows that the change in excess heat in response to a temperature change corresponds to the same change in the quasipotential, or in other words\begin{equation}\label{eh}
\delta Q^\text{exc} =   - \langle \id V\rangle,
\end{equation}
where we left out the subscripts. It is important to see that under detailed balance, $q_\beta(x,y) = E(x) - E(y)$ and
$\delta Q^\text{eq} = \id \langle E\rangle^\text{eq} - \langle \id E\rangle^\text{eq}$ as in the First Law. Since  $\langle V\rangle =0$, we have $-\langle \,\id V\,\rangle
= \sum_{y} \id\rho(y)\,V(y)$ where $\id\rho(y)$ is the change of the (unique) stationary distribution $\rho$ under $\beta\rightarrow \beta+\id \beta$.  In other words, 
\begin{equation}\label{eha}
\delta Q^\text{exc} =  \sum_{y} \id\rho(y)\,V(y)= \langle V\,;\,\id\log \rho\rangle,
\end{equation}
where $\langle f\,;\,g\rangle = \langle fg\rangle - \langle f\rangle\langle g\rangle$ denotes the stationary covariance, and we used that $\langle \id \log\rho\rangle=0$.  Note that in equilibrium, $V=E-\langle E\rangle$ and $\log \rho =-\beta E - \log Z$ for the partition function $Z$, and the excess heat (the total heat) gets expressed in terms of the energy-variance, which assures its positivity. In nonequilibrium regimes, we get two `potentials' (formally, $V$ and $\log \rho$), and their structure and mutual correlation shape the excess heat following the response relation \eqref{eha}.\\

The concept of excess heat naturally leads to the definition of nonequilibrium heat capacity. The foundational theory and initial examples of nonequilibrium heat capacities can be found in \cite{eu, jir}. Nonequilibrium heat capacity is
\begin{equation}\label{ct}
C(T) =  - \Big\langle \frac{\partial}{\partial T} V \Big\rangle_T,\quad \text{or}\quad C(\beta)=\beta^2 \Big\langle \frac{\partial}{\partial \beta } V \Big\rangle_\beta,
\end{equation}
which measures the excess heat towards the system per temperature ($=\beta$ in that case) change. The starting point is simply \eqref{eh}. Still, the change in $\beta$ corresponds to a temperature shift, denoted as $T\rightarrow T' = T +\id T$, while keeping other controls constant, such as energy levels (similar to fixed volume specific heats).\\

The quasipotential \eqref{psp} and the heat capacity \eqref{ct} are calculated in papers \cite{aaron, pritha1, pritha2}, where different aspects of nonequilibrium calorimetry are explored. The quasipotential $V$ can also be expressed as the solution of the Poisson equation $LV=\langle \mathcal{P}\rangle-\mathcal{P}$. The graphical representation of $V$  is given in the paper \cite{pois}, which is utilized for calculating heat capacity in various physical models, as discussed in \cite{irene, simon}. 

\section{Nonequilibrium Third Law of thermodynamics }\label{3rdlaw}
The Third Law asserts that the excess heat, as defined for quasistatic processes between nonequilibrium conditions, vanishes at absolute zero:
\begin{equation}\label{3rd}
\delta Q^\text{exc} \rightarrow 0.
\end{equation}
Consequently, the heat capacity must also vanish at zero temperature, $C(T\downarrow 0)\rightarrow 0$. The right-hand side in \eqref{eha} depends on the temperature $T$ via $V$ and $\rho$. In order to show  \eqref{3rd}, two things suffice: 1)  $\id \rho\rightarrow 0$ for $T\downarrow 0$ and 2) bounded $V$.\\
Suppose a unique state $x^*$ exists such that the stationary distribution $\rho(x)$ converges to $\delta_{x,x^*}$ (Kronecker delta) as $T$ approaches zero. Additionally, assume the smoothness of $\rho$, allowing the interchangeability of the limit $T$ with the derivative with respect to $\beta$ (or $T$). Considering both assumptions, $\id \rho$ goes to zero in the limit of $T\downarrow 0$. If the dominant state is unique, the convergence of $\id \rho$ at low temperatures is exponentially fast. However, if there are multiple dominant states, $\id \rho\rightarrow 0$ may exhibit polynomial behaviour with respect to temperature. The way the stationary condition $\rho$ influences the excess heat at low temperatures is a noteworthy example of how heat, even in nonequilibrium statistical mechanics, reveals insights into low-temperature degeneracy. \\

 Secondly, from \eqref{psp}, it is evident that at low temperatures, the boundedness of $V$ is linked to relaxation properties and the accessibility of states. To show the boundedness of  $V(x)$ it is enough, we show that for every edge $e=(x,y),$ the differences of quasipotentials $V(e) = V_{\beta'}(x)-V_{\beta'}(y)$ is bounded \cite{SM}. The subsequent section will demonstrate that the differences in quasipotentials over an edge are bounded only when the mean time to reach any other states for the first time from the edge is also bounded. Notably, under global detailed, balanced condition,  if in \eqref{ju} $q(x,y) = E(x) - E(y),$ then $V(x) = E(x) - \langle E\rangle,$ and the boundedness is immediately guaranteed.
 In \cite{SM}, the boundedness of the quasipotential is investigated through graphical representation, where the second condition is precisely studied. It is worth mentioning that in the limit of close to equilibrium, the quasipotential is always bounded, and the approximation to the nonequilibrium heat capacity always goes to zero as temperature goes to zero, see \cite{mcheat}. 
 
\section{Bound for quasipotential in terms of the mean first-passage times}
Let $\tau(x, y)$  represent the mean time for a random walker starting from $x$ to reach $y$ for the first time. In \cite{pois}, the graphical representations of the quasipotential and the mean first passage time within the framework of Markov jump processes are provided, where $\tau$ satisfies the Poisson equation $L \tau=-1$. In the same paper, the graphical representations of the  quasipotential and the mean first passage time provide the following relation:
\begin{align}\label{difv}
  V(x)-V(y)=\sum_u \, \rho(u) \, f(u)[\tau(y,u) -\tau(x,u )],
\end{align}
where $f=  \cal P - \langle \cal P\rangle$. As a result, the difference in quasipotentials over an edge is bounded if the mean times to reach any other state, for the first time, from the edge are bounded. In simpler terms, the absence of local traps in the graph ensures this bound. This interpretation again leads to the second condition of the Third Law, which pertains to the accessibility of all states in the graph. \\
In the same paper, an upper bound is given to the difference of quasipotentials, put  $||f||:= \max_{z\in K}|f(z)|$,
 and the bound
    \begin{equation}\label{bb}
    |   V(x) - V(y)| \leq ||f||\min\{\tau(x,y),\tau(y,x)\}.
    \end{equation}
    The bound \eqref{bb} is not always optimal. It can still be improved as given in \cite{pois}.

\section{Conclusions }
Excess heat in nonequilibrium systems, representing absorbed heat during quasistatic relaxation, defines nonequilibrium heat capacity. Considering Markov jump processes on a connected graph, we express excess heat in the quasistatic relaxation to a new nonequilibrium condition. The excess heat is the product of  $\id \rho$ in stationary probabilities and quasipotential $V$. To establish our main theorem, the boundedness of the quasipotential in low temperature is a prerequisite. In the last section, we provide an expression of quasipotential in terms of mean first passage time, offering an interpretation of the accessibility of states. Consequently,  in the low temperature, $T\downarrow0$, the quasipotential remains bounded if the states are accessible enough.   \textit{Our main result asserts that the nonequilibrium heat capacity tends to zero at low temperatures, contingent on two conditions: the unique ground state and a bounded quasipotential.}\\

\textbf{Acknowledgments: } The Author is  grateful to C. Maes and K. Net\^ocn\'y  for  insightful discussions during the  collaborative papers \cite{nernst, SM, pois}. Special thanks to C. M. for his meticulous review of the manuscript.
\section*{References}
\bibliographystyle{iopart-num}
\bibliography{BibTeX/iopart-num}

\end{document}